\documentclass[11pt,letterpaper]{article}

\usepackage[portrait,margin=2.5cm]{geometry}

\usepackage{geometry}
\usepackage{ upgreek }
\usepackage{setspace}
\usepackage{titlesec}
\usepackage[titles]{tocloft}
\usepackage{epsfig}
\usepackage{epstopdf}
\usepackage{multirow}
\usepackage[font=small,labelfont=bf]{caption}
\usepackage{enumerate}
\usepackage{dsfont}
\usepackage{tikz}
\usepackage{pgfplots}
\usepackage{amssymb}
\usepackage{subcaption}
\usepackage[ruled,vlined]{algorithm2e}

\usepackage[round]{natbib} 

\usepackage[font=small,skip=-4pt]{subcaption}
\usepackage{color}

\usepackage{amsthm}
\usepackage{amsmath}
\usepackage{amssymb}

\usepackage{xspace}

\usepackage[hidelinks,linktocpage=true]{hyperref} 

\newcounter{Appendix}

\renewcommand{\cite}{\citep}

\theoremstyle{plain}

\newtheoremstyle{mydef}
        {3pt}
        {3pt}
        {\normalfont}
        {}
        {\bfseries}
        {.}
        {.5em}
        {\thmname{#1} \thmnumber{#2}\thmnote{#3}}

\newcommand\rightparend[1]{{%
      \unskip\nobreak\hfil\penalty50
      \hskip2em\hbox{}\nobreak\hfil\textbf{#1}%
      \parfillskip=0pt \finalhyphendemerits=0 \par}}

\newtheorem{xxexample}{Example}[section]

\usepackage{epsfig,amssymb,amsmath, multirow, url, amsfonts, tcolorbox,booktabs, enumitem}

\usepackage{amsthm}
\usepackage{natbib}
\usepackage[utf8]{inputenc}
\usepackage{bbm}
\usepackage{dsfont}

\usepackage{hyperref}
\usepackage{graphicx}
\usepackage{upref,setspace}
\usepackage{caption}

\usepackage{enumerate}

\numberwithin{equation}{section}
\numberwithin{figure}{section}
\numberwithin{table}{section}

\newtheorem{thm}{Theorem}[section]
\newtheorem{lem}[thm]{Lemma}
\newtheorem{cor}[thm]{Corollary}

\newtheorem*{ass*}{Assumption}

\theoremstyle{definition}
\newtheorem{rem}[thm]{Remark}

\renewcommand{\leq}{\leqslant}
\renewcommand{\geq}{\geqslant}

\newcommand{\wt}{\widetilde}
\newcommand{\wh}{\widehat}

  \let\gc=\gamma  \let\gee=\epsilon
 \let\gh=\eta            \let\go=\omega   \let\gs=\sigma  
  \let\gz=\zeta
\let\gC=\Gamma  \let\gF=\Phi  
       \let\gPs=\Psi    

\newcommand{\cA}{\mathcal{A}}

\newcommand{\cL}{\mathcal{L}}
\newcommand{\cM}{\mathcal{M}}
\newcommand{\cP}{\mathcal{P}}

\newcommand{\mv}[1]{\boldsymbol{#1}}

\newcommand{\mvA}{\boldsymbol{A}}\newcommand{\mvB}{\boldsymbol{B}}\newcommand{\mvC}{\boldsymbol{C}}
\newcommand{\mvD}{\boldsymbol{D}}
\newcommand{\mvI}{\boldsymbol{I}}

\newcommand{\mvM}{\boldsymbol{M}}\newcommand{\mvN}{\boldsymbol{N}}
\newcommand{\mvQ}{\boldsymbol{Q}}\newcommand{\mvR}{\boldsymbol{R}}
\newcommand{\mvS}{\boldsymbol{S}}\newcommand{\mvU}{\boldsymbol{U}}
\newcommand{\mvV}{\boldsymbol{V}}\newcommand{\mvW}{\boldsymbol{W}}\newcommand{\mvX}{\boldsymbol{X}}
\newcommand{\mvY}{\boldsymbol{Y}}\newcommand{\mvZ}{\boldsymbol{Z}}

\newcommand{\mve}{\boldsymbol{e}}\newcommand{\mvf}{\boldsymbol{f}}

\newcommand{\mvga}{\boldsymbol{\alpha}}

\newcommand{\mvgee}{\boldsymbol{\epsilon}}
\newcommand{\mvgF}{\boldsymbol{\Phi}}\newcommand{\mvgTh}{\boldsymbol{\Theta}}
\newcommand{\mvgh}{\boldsymbol{\eta}}
\newcommand{\mvgL}{\boldsymbol{\Lambda}}

\newcommand{\mvgS}{\boldsymbol{\Sigma}}

\newcommand{\mvgz}{\boldsymbol{\zeta}}

\newcommand{\bC}{\mathbb{C}}

\newcommand{\bN}{\mathbb{N}}
\newcommand{\bR}{\mathbb{R}}

\newcommand{\bZ}{\mathbb{Z}}

\DeclareMathOperator{\E}{\mathds{E}}

\usepackage{amsmath}

\newtheorem{@assumption}{\sc Assumption}

\newcommand{\wtmvX}{\widetilde{\mvX}}

\newcommand{\wtmvgee}{\widetilde{\mvgee}}

\newcommand{\wtmvgF}{\widetilde{\mvgF}}

\newcommand{\wbmvgL}{\bar{\mvgL}}
\newcommand{\wbmvf}{\bar{\mvf}}
\newcommand{\wbmvgF}{\bar{\mvgF}}

\newcommand{\wb}{\bar}

\begin{document}
\sloppy
\title{Dynamic factor and VARMA models: equivalent representations, dimension reduction and nonlinear matrix equations
\footnote{The first and third authors were supported in part by the NSF grants DMS-2113662 and DMS-2134107. The second author were supported in part by NSF grant DMS-2134107.}}

\author{Shankar Bhamidi \quad \; Dhruv Patel \quad \;  Vladas Pipiras \\ 
University of North Carolina}

\date{\today}
\maketitle

\abstract{A dynamic factor model with factor series following a VAR$(p)$ model is shown to have a VARMA$(p,p)$ model representation. Reduced-rank structures are identified for the VAR and VMA components of the resulting VARMA model. It is also shown how the VMA component parameters can be computed numerically from the original model parameters via the innovations algorithm, and connections of this approach to non-linear matrix equations are made. Some VAR models related to the resulting VARMA model are also discussed. } 

\section{Introduction} 

Let $\mvX_t$, $t \in \bZ$, be a $d$-vector time series following a dynamic factor model (DFM) in its static form given by 
\begin{equation}\label{DFM}
\mvX_t = \mvgL \mvf_t + \mvgee_t,
\end{equation}
where $\{\mvf_t\}$ is a $r$-vector stationary time series, $\mvgL$ is a $d\times r$ loadings matrix and $\mvgee_t$ are the error terms with $\E \mvgee_t = 0$, $\E \mvgee_t \mvgee_s = \mv0$ for $t\neq s$ and $\E \mvgee_t \mvgee_t' = \mvgS_{\gee}$, i.e., $\{\mvgee_t\} \sim $ WN$(\mv0, \mvgS_{\gee})$ is a white noise series. The factors $\{\mvf_t\}$ can be correlated across time and are assumed to have zero mean. The errors $\{\mvgee_t\}$ and factors $\{\mvf_t\}$ are supposed to be uncorrelated. We assume further that the factors follow a stationary vector auto-regressive (VAR) model of order $p \in \bN_0 := \{0,1,...\}$, VAR$(p)$, namely,
\begin{equation}\label{VARfactors}
\mvf_t = \sum_{i=1}^p \mvgF_i \mvf_{t-i} + \mvgh_t,
\end{equation}
where $\mvgF_i$, $i = 1,\ldots,p$, are $r\times r$ VAR coefficient matrices and $\{\mvgh_t\} \sim $ WN$(\mv0, \mvgS_{\gh})$. It is the factor series $\{\mvf_t\}$ which drives the temporal dynamics of the series $\{\mvX_t\}$. The assumed VAR structure for the factors is flexible from the modeling perspective. Owing to their importance in multiple disciplines, DFMs have been studied extensively, see e.g.\ \citet{bai:2008}, \citet{stock:2011}, and \citet{doz:2011} and the references therein. 

It is part of the folklore in time series research that the DFM \eqref{DFM}--\eqref{VARfactors} can be rewritten as a Vector AutoRegressive Moving Average (VARMA) model. An informal argument for this is given in Section \ref{s-rewriteDFM} below. A VARMA model for $\mvX_t \in \bR^{d}$ of orders $p,q \in \bN_0$, denoted VARMA$(p,q)$, is given by 
\begin{equation}\label{VARMA}
\mvX_t = \sum_{i=1}^p \wtmvgF_i \mvX_{t-i} + \mvgz_t + \sum_{j=1}^q \mvgTh_j \mvgz_{t-j},
\end{equation}
where $\{\mvgz_t\} \sim $ WN$(\mv0 , \mvgS_{\gz}) $. The matrices $\wtmvgF_i \in \bR^{d\times d}$ and $\mvgTh_j \in \bR^{d\times d}$ are referred to as the vector autoregressive (VAR) coefficient matrix of order $i$ and the vector moving average (VMA) coefficient matrix of order $j$, respectively. The series $\{\mvX_t\}$ follows a vector moving average model of order $q$, VMA$(q)$, if $p=0$. An infinite VMA representation will refer to the case when $q=\infty$.

In this paper, we show that the DFM \eqref{DFM}--\eqref{VARfactors} can in fact be rewritten as a VARMA$(p,p)$ model. Identifying the VAR component of the VARMA$(p,p)$ model is straightforward (Section \ref{s-rewriteDFM} below) but dealing with the VMA component is more delicate. Not only do we establish the existence of the VMA component, we also show that it has a reduced-rank structure; the same holds for the VAR component. Furthermore, we show that the VMA matrices can be computed numerically from the parameters of the DFM \eqref{DFM}--\eqref{VARfactors} by the innovations algorithm and we make interesting connections of this approach to non-linear matrix equations.

As a result, this work allows one to use theory and techniques built for VARMA models on the DFM. For example, the best linear predictor of the DFM \eqref{DFM}--\eqref{VARfactors} can be obtained using the VARMA representation as shown in Section \ref{s-forecasting}. Additionally, the computational complexity of the best linear predictor can be significantly reduced by leveraging the reduced-rank structure of the DFM. Finally, we draw contrasts and parallels of the established VARMA$(p,p)$ model to other related models.

VARMA and DFM are two common approaches to modeling multivariate, possibly high-dimensional time series. They often seem to be viewed and treated somewhat separately, with VARMA based on regression and DFM connected to dimension reduction. Our work clarifies connections, and differences, between these two fundamental modeling approaches.

The rest of this paper is structured as follows. Preliminary observations can be found in Section \ref{s-prelim}, including candidate VAR and VMA components of the VARMA$(p,p)$ model. The dimension reduction and existence of the VMA component are considered in Sections \ref{s-dim_red} and \ref{s-VMA_rep}, respectively. Section \ref{s-VMA_rep} further includes connections to the innovations algorithm and non-linear matrix equations. Section \ref{s-main_res} contains our main result, some implications, and connections to related models. Section \ref{s-conclusion} concludes the paper. Proofs are deferred to the Appendix.

\section{Preliminaries}\label{s-prelim}

\subsection{Simplified Form of DFM}\label{s-simplify_DFM}
In this section, we make some assumptions which simplify the DFM \eqref{DFM}--\eqref{VARfactors} but do not impose restrictions on the model. 
We can assume without loss of generality for the purposes here that 
\begin{equation}\label{error cov}
\mvgS_{\gee} = \mvI_d.
\end{equation}
Indeed, setting
\begin{equation}
\wtmvX_t := \mvgS_{\gee}^{-\frac{1}{2}}\mvX_t = \mvgS_{\gee}^{-\frac{1}{2}} \mvgL \mvf_t + \mvgS_{\gee}^{-\frac{1}{2}} \mvgee_t = \wt\mvgL \mvf_t + \wt\mvgee_t,
\end{equation}
where $\{\wtmvgee_t\} \sim $ WN$(\mv0,\mvI_d)$, if $\wtmvX_t$ can be written as a VARMA$(p,p)$ with coefficient matrices $\wt\mvgF_i^*, \mvgTh_j^*,$ and $\mvgS_{\gz}^*$, then so can $\mvX_t = \mvgS_{\gee}^{\frac{1}{2}} \wtmvX_t$ with 
\begin{equation}\label{PhiTilde_noniden_errorcov}
\wt\mvgF_i = \mvgS_{\gee}^{\frac{1}{2}} \wt\mvgF_i^* \mvgS_{\gee}^{-\frac{1}{2}}, \; \mvgTh_j = \mvgS_{\gee}^{\frac{1}{2}} \mvgTh_j^* \mvgS_{\gee}^{-\frac{1}{2}}, \; \text{and } \mvgS_{\gz} =  \mvgS_{\gee}^{\frac{1}{2}} \mvgS^*_{\gz} \mvgS_{\gee}^{\frac{1}{2}}.
\end{equation} 

Furthermore, since $\mvgL'\mvgL$ is a symmetric matrix, there exist orthogonal $\mvS$ and diagonal $\mvD$ such that $\mvgL'\mvgL = \mvS \mvD \mvS'$. Define $\wbmvgL = d^{\frac{1}{2}}\mvgL \mvS \mvD^{-\frac{1}{2}}$ and $\wbmvf_t = d^{-\frac{1}{2}}\mvD^{\frac{1}{2}} \mvS' \mvf_t$. Then, $\mvX_t$ follows a DFM given by

\begin{equation}\label{simplifiedX}
\mvX_t = \wbmvgL \wbmvf_t + \mvgee_t
\end{equation}
with $\wbmvgL'\wbmvgL =  d\mvD^{-\frac{1}{2}}\mvS'\mvgL'\mvgL \mvS \mvD^{-\frac{1}{2}} = d\mvI_r$, and 
\begin{equation}\label{rewrite-fbar}
\wbmvf_t = d^{-\frac{1}{2}}\mvD^{\frac{1}{2}} \mvS' \mvf_t = \sum_{i=1}^p d^{-\frac{1}{2}}\mvD^{\frac{1}{2}} \mvS' \mvgF_i \mvS \mvD^{-\frac{1}{2}}\mvD^{\frac{1}{2}} \mvS' \mvf_{t-1} + d^{-\frac{1}{2}}\mvD^{\frac{1}{2}} \mvS' \mvgh_t= \sum_{i=1}^p \wbmvgF_i \wbmvf_{t-i} + \wb\mvgh_t, 
\end{equation}
where $\wbmvgF_i = \mvD^{\frac{1}{2}} \mvS' \mvgF_i \mvS \mvD^{-\frac{1}{2}}$ for $i = 1,\dots,p$, and $\wb\mvgh_t = d^{-\frac{1}{2}}\mvD^{\frac{1}{2}} \mvS' \mvgh_t$. Thus, we may assume without loss of generality in \eqref{DFM} that 
\begin{equation}\label{lambda identity}
\mvgL'\mvgL = d\mvI_r.
\end{equation}

The dimension $d$ appears in \eqref{lambda identity} in order to make explicit the dependence of $d$ in certain quantities below; the expression \eqref{lambda identity} is also consistent with the standard assumption on the loadings (e.g.\ \citet{stock:2011} and \citet{doz:2011}) in the setting of ``strong" factors, where the eigenvalues of $\mvgL'\mvgL$ are of order $d$. For the rest of paper we will assume that the DFM \eqref{DFM}--\eqref{VARfactors} satisfies \eqref{error cov} and \eqref{lambda identity}. 

\subsection{Candidate VAR and VMA Components of DFM }\label{s-rewriteDFM}

Our goal is to show that a time series $\{\mvX_t\}$ satisfying a DFM \eqref{DFM}--\eqref{VARfactors}, can be represented as VARMA$(p,p)$ in \eqref{VARMA}, for some $\wtmvgF_i, \mvgTh_i, \mvgS_{\gz}$. We do so by rewriting the DFM into the following candidate VAR and VMA components.

Using \eqref{lambda identity} and \eqref{DFM}, we have
\begin{equation}
\mvgL' \mvX_t = d \mvf_t + \mvgL'\mvgee_t
\end{equation}
and thus,
\begin{align}
\mvX_t &= \mvgL \mvf_t + \mvgee_t \nonumber \\
	   &= \sum_{i=1}^p \mvgL \mvgF_i \mvf_{t-i} + \mvgL \mvgh_t + \mvgee_t \nonumber \\
	   &= \sum_{i=1}^p \frac{1}{d}\mvgL \mvgF_i \mvgL' \mvX_{t-i}- \sum_{i=1}^p \frac{1}{d}\mvgL \mvgF_i \mvgL' \mvgee_{t-i} + \mvgL \mvgh_t + \mvgee_t  \nonumber\\ 
	   &= \sum_{i=1}^p \wtmvgF_i \mvX_{t-i} + \mvgL \mvgh_t + \mvgee_t - \sum_{i=1}^p \wtmvgF_i \mvgee_{t-i}, \label{rewriteDFM}
\end{align}
where 
\begin{equation}\label{def-Phi_tilde}
\wtmvgF_i = \frac{1}{d}\mvgL \mvgF_i \mvgL'
\end{equation} for $i=1,\ldots, p$. The term $ \sum_{i=1}^p \wtmvgF_i \mvX_{t-i}$ is the VAR component. We wish to show that the remaining terms in \eqref{rewriteDFM}, namely,
\begin{align}\label{def-Z}
\mvZ_t := \mvgL \mvgh_t + \mvgee_t - \sum_{i=1}^p \wtmvgF_i \mvgee_{t-i},
\end{align}
can be represented as a VMA$(p)$ process
\begin{equation}\label{VMA_Z}
\mvZ_t  = \mvgz_t +  \sum_{i=1}^p \mvgTh_i \mvgz_{t-i},
\end{equation}
for some $\{\mvgz_t\} \sim$ WN$(\mv0,\mvgS_{\gz})$ and $\mvgTh_i \in \bR^{d\times d}$ for $i = 1,\ldots,p$. Since $\{\mvgh_t\} \sim$ WN$(\mv0,\mvgS_{\gh})$, $\{\mvgee_t\} \sim$ WN$(\mv0,\mvI_d)$, and these errors are assumed to be uncorrelated, the autocovariance function (ACVF) of the process $\{\mvZ_t\}$ is given by 
\begin{equation}\label{ACVF_Z}
\Gamma_Z(h) := \E \mvZ_{t+h}\mvZ_t' = 
    \begin{cases}
        \mvgL\mvgS_{\gh}\mvgL' + \mvI_d +  \sum_{i=1}^p\wtmvgF_i \wtmvgF_i', & \text{if } h = 0,\\
        \sum_{i=0}^{p-h} \wtmvgF_{i+h} \wtmvgF_{i}', & \text{if } h = 1,\ldots,p, \\
        \mv0, & \text{if } h \geq p+1,
    \end{cases}
\end{equation}
where $\wt\mvgF_0 := - \mvI_d$ and we used \eqref{error cov}. As the ACVF \eqref{ACVF_Z} is $\mv0$ for $|h| \geq p+1$, it is natural to expect that $\{\mvZ_t\}$ has a VMA$(p)$ representation. Equating the ACVFs of \eqref{def-Z} and \eqref{VMA_Z} at lags $0,1,\ldots, p$ leads to a system of $p+1$ equations given by 
\begin{align}
\gC_Z(0) &= \mvgL\mvgS_{\gh}\mvgL' + \mvI_d +  \sum_{i=1}^p\wtmvgF_i \wtmvgF_i' = \mvgS_{\gz} +  \sum_{i=1}^p\mvgTh_i \mvgS_{\gz} \mvgTh_i', \label{full dim Lag0}\\
\gC_Z(h) &= \sum_{i=0}^{p-h} \wtmvgF_{i+h} \wtmvgF_{i}'  = \sum_{i=0}^{p-h} \mvgTh_{i+h} \mvgS_{\gz} \mvgTh_{i}', \quad \quad h = 1,\ldots,p,\label{full dim Lag1-p}
\end{align}
where $-\wt\mvgF_0 = \mvgTh_0 = \mvI_d$. The issue is that, to the best of our knowledge, it is not known whether the system \eqref{full dim Lag0}--\eqref{full dim Lag1-p} has a solution in terms of $\mvgTh_i,\mvgS_{\gz}$, or how to compute a solution (even numerically). This issue arises for dimension $d>1$ due to the non-commutativity of various matrix operations; when $d=1$, the system \eqref{full dim Lag0}--\eqref{full dim Lag1-p} reduces to a quadratic equation that can be solved explicitly. In Section \ref{s-VMA_rep}, we show that, for general $d$, a solution exists and provide a method to numerically compute a solution.

\section{Dimension Reduction of VMA Component}\label{s-dim_red}
Another issue with \eqref{full dim Lag0}--\eqref{full dim Lag1-p}, even from a numerical standpoint, is the setting of large dimension $d$, as DFMs arise often in modeling high-dimensional time series. Note that the VAR transition matrices in \eqref{def-Phi_tilde} for the VARMA model have reduced rank. We show in this section that if a solution to \eqref{full dim Lag0}--\eqref{full dim Lag1-p} exists, then it must have a reduced-rank structure. By considering a special case in Section \ref{s-one_factor_case}, we are able to guess and then prove the reduced-rank structure in the general case in Section \ref{s-general_case}.

\subsection{Case of One Factor}\label{s-one_factor_case}
Consider the case when $r = 1$ and $p=1$. Then, the system \eqref{full dim Lag0}--\eqref{full dim Lag1-p} results in two equations in $\mvgTh_1$ and $\mvgS_{\gz}$,
\begin{align}
\mvgL \mvgS_{\gh} \mvgL' + \mvI_d + \frac{1}{d}\mvgL\mvgF_1\mvgF_1'\mvgL' &= \mvgS_{\gz} + \mvgTh_1\mvgS_{\gz}\mvgTh_1', \label{rewriteDFM-Lag0} \\ 
-\frac{1}{d}\mvgL \mvgF_1\mvgL' & = \mvgTh_1\mvgS_{\gz}, \label{rewriteDFM-Lag1} 
\end{align}
where we used \eqref{def-Phi_tilde}.
If $r=1$, then $\mvgL \in \bR^{d}$, and \eqref{lambda identity} implies that $\mvgL'\mvgL = \|\mvgL\|_2^2 = d$. Also, write $\mvgF_1 = \gF_1 \in \bR$ and $\mvgS_{\gh} = \gs^2_{\gh}$. Then, \eqref{rewriteDFM-Lag0} and \eqref{rewriteDFM-Lag1} are expressed as 
\begin{align}
\gs^2_{\gh} \mvgL\mvgL' + \mvI_d + \frac{\gF_1^2}{d} \mvgL\mvgL' & = \mvgS_{\gz} + \mvgTh_1\mvgS_{\gz}\mvgTh_1', \label{motivation-Lag0} \\
-\frac{\gF_1}{d} \mvgL\mvgL' & = \mvgTh_1\mvgS_{\gz} \label{motivation-Lag1}.
\end{align}
Assuming $\mvgS_{\gz}$ is positive definite (invertible), it follows that
\begin{equation}\label{Theta1_onefactor}
\mvgTh_1 = -\frac{\gF_1}{d}\mvgL \mvgL' \mvgS_{\gz}^{-1}
\end{equation}
and hence
\begin{equation}\label{theta1Sigmatheta1_onefactor}
\mvgTh_1\mvgS_{\gz}\mvgTh_1' = \frac{\gF_1^2}{d^2} \mvgL\mvgL'\mvgS_{\gz}^{-1}\mvgL \mvgL' = \frac{\gF_1^2}{d^2} (\mvgL'\mvgS_{\gz}^{-1}\mvgL )\mvgL\mvgL',
\end{equation}
since $\mvgL'\mvgS_{\gz}^{-1}\mvgL  \in \bR$. 

Then, by \eqref{theta1Sigmatheta1_onefactor}, \eqref{motivation-Lag0} can be rewritten as 
\begin{equation}
\mvI_d + \bigg (\gs^2_{\gh} +\frac{\gF_1^2}{d^2} \bigg ) \mvgL\mvgL' = \mvgS_{\gz} + \frac{\gF_1^2}{d^2} (\mvgL'\mvgS_{\gz}^{-1}\mvgL )\mvgL\mvgL' = \mvgS_{\gz} + \frac{\gF_1^2}{d^2} \go \mvgL\mvgL' ,
\end{equation} 
where $\go = \mvgL'\mvgS_{\gz}^{-1}\mvgL$, and hence
\begin{equation}
\mvgS_{\gz} = \mvI_d + \bigg (\gs^2_{\gh} + \frac{\gF_1^2}{d^2}- \frac{\gF_1^2}{d^2} \go \bigg ) \mvgL\mvgL' =: \mvI_d + u \mvgL'\mvgL.
\end{equation}
Setting $ C = \frac{u}{1 + ud}$, $\mvgS_{\gz}^{-1} = \mvI_d  - C \mvgL\mvgL'$ and, by \eqref{Theta1_onefactor},
\begin{equation}
\begin{split}
\mvgTh_1 & = -\frac{\gF_1}{d} \mvgL \mvgL' \mvgS_{\gz}^{-1} \\
& = -\frac{\gF_1}{d} \mvgL \mvgL' \bigg (\mvI_d  - C \mvgL\mvgL' \bigg ) \\
& = \bigg (-\frac{\gF_1}{d} + \gF_1 C \bigg ) \mvgL\mvgL' =: v\mvgL'\mvgL.
\end{split}
\end{equation}
So, there exist some constants $u,v \in \bR$ such that  
\begin{align}
\mvgS_{\gz} & = \mvI_d + u\mvgL\mvgL', \label{motivation-Sigma} \\
\mvgTh_1 &= v\mvgL \mvgL' \label{motivation-Theta}.
\end{align}
By substituting \eqref{motivation-Sigma}--\eqref{motivation-Theta} back into \eqref{motivation-Lag0}--\eqref{motivation-Lag1}, the latter system becomes
\begin{align}
\mvI_d + \bigg (\gs^2_{\gh} + \frac{\gF_1^2}{d}\bigg ) \mvgL\mvgL' & = \mvI_d + u\mvgL\mvgL' + v\mvgL \mvgL'(\mvI_d + u\mvgL\mvgL')v\mvgL \mvgL', \\
-\frac{\gF_1}{d} \mvgL\mvgL' & = v\mvgL \mvgL'(\mvI_d + u\mvgL\mvgL')
\end{align}
and is satisfied when 
\begin{equation}\label{onedim-sol}
\gs^2_{\gh} + \frac{\gF_1^2}{d} = u + dv^2(1+du), \; -\frac{\gF_1}{d} = v(1+du),
\end{equation}
which can be reduced to solving a quadratic equation in $u$. The discriminant of this quadratic is given by $(1 - d\gs^2_\eta - \gF_1^2)^2 + 4d\gs^2_\eta$ which is greater than or equal to zero.

\subsection{General Case}\label{s-general_case}

Motivated by \eqref{motivation-Sigma}--\eqref{motivation-Theta}, we start our analysis by postulating that the solutions (if they exist) $\mvgS_{\gz}$ and $\mvgTh_i$, $i=1,\ldots, p$, solving \eqref{full dim Lag0}--\eqref{full dim Lag1-p} can be written as 
\begin{align}
\mvgS_{\gz} & = \mvI_d + \frac{1}{d} \mvgL \mvU \mvgL', \label{lowdim-Sigma} \\
\mvgTh_i &= \frac{1}{d} \mvgL \mvV_i \mvgL', \quad i=1,\ldots,p, \label{lowdim-Theta}
\end{align}
with $\mvU,\mvV_i \in \bR^{r\times r}$ and $\mvI_r + \mvU$ symmetric positive definite, since \eqref{lowdim-Sigma} implies $\mvI_r + \mvU = \frac{1}{d}\mvgL'\mvgS_{\gz}\mvgL$.

Substituting \eqref{lowdim-Sigma}--\eqref{lowdim-Theta} into \eqref{full dim Lag0}--\eqref{full dim Lag1-p}, using \eqref{lambda identity}, and removing $\mvgL,\mvgL'$ from the two sides of the expressions, we obtain that \eqref{full dim Lag0}--\eqref{full dim Lag1-p} is satisfied if the following system of matrix equations in dimension $r$ is satisfied in terms of $\mvV_i,\mvU$:
\begin{align}
\mvI_r + d\mvgS_{\gh} + \sum_{i=1}^p \mvgF_i \mvgF_i' &= \mvI_r + \mvU + \sum_{i=1}^p \mvV_i (\mvI_r + \mvU) \mvV_i', \label{lowdim-MVLag0}\\
\sum_{i=0}^{p-h} \mvgF_{i+h} \mvgF_{i}' &= \sum_{i=0}^{p-h}  \mvV_{i+h} (\mvI_r + \mvU) \mvV_{i}', \quad h=1,\ldots,p,\label{lowdim-MVLag1-p}
\end{align}
where $\mvV_0 = -\mvgF_0 = \mvI_r$. Note that the left-hand sides of \eqref{lowdim-MVLag0}--\eqref{lowdim-MVLag1-p} are the autocovariances of the reduced-dimension $r$-vector series
\begin{equation}\label{def-reducedrank_Z}
\frac{1}{\sqrt{d}}\mvgL' \mvZ_t = \sqrt{d} \mvgh_t + \frac{1}{\sqrt{d}}\mvgL' \mvgee_t - \sum_{i = 1}^p \frac{1}{\sqrt{d}} \mvgF_i \mvgL ' \mvgee_{t-i}.
\end{equation} 
Then the following theorem holds, with the proof found in Appendix \ref{appendix-thm3.1}.
\begin{thm}\label{thm-VMAdimred}
If $\{\frac{1}{\sqrt{d}}\mvgL' \mvZ_t\}$ is a VMA$(p)$ process with error covariance $\mvI_r + \mvU$ and VMA matrices $\mvV_1,\ldots,\mvV_p$, then $\{\mvZ_t\}$ is a VMA$(p)$ series with error covariance and VMA matrices given by \eqref{lowdim-Sigma}--\eqref{lowdim-Theta}.
\end{thm}
In fact, it will be shown in Section \ref{s-VMA_rep} below that $\{\frac{1}{\sqrt{d}}\mvgL' \mvZ_t\}$ does have a VMA$(p)$ representation. So by Theorem \ref{thm-VMAdimred}, $\{\mvZ_t\}$ also admits a VMA$(p)$ representation with the quantities \eqref{lowdim-Sigma}--\eqref{lowdim-Theta} having a reduced-rank structure. We also note that the matrix $\mvU$ need not be positive definite, but $\mvI_r + \mvU$, the covariance matrix, must be. Indeed, we observe numerically that $\mvU$ can have negative eigenvalues in $(-1,0)$. Furthermore, when $p = 0$, by \eqref{lowdim-MVLag0} we get that $d\mvgS_{\gh} = \mvU$, so $\mvU$ is of order $d$. For $p\geq 1$, we still expect $\mvU$ to be of order $d$ (and see this in numerical experiments) and $\mvV_i$, $i = 1,\ldots, p$, to be of order $\frac{1}{d}$ because of the term $\mvV_h (\mvI_r + \mvU) \mvV_0' = \mvV_h (\mvI_r + \mvU)$ with $i = 0$ on the right-hand side of \eqref{lowdim-MVLag1-p}, that is, 
\begin{equation}\label{UV-orders}
\mvU \asymp d \;, \quad \mvV_i \asymp \dfrac{1}{d} \;, \; i = 1,\ldots,p.
\end{equation}
The orders in \eqref{UV-orders} should also not be surprising from the following perspective. The order of $\mvU$ is determined by the term $\sqrt{d}\mvgh_t$ in \eqref{def-reducedrank_Z}. On the other hand, by the functional central limit theorem, we expect the term $\frac{1}{\sqrt{d}}\mvgL' \mvgee_t - \sum_{i = 1}^p \frac{1}{\sqrt{d}} \mvgF_i \mvgL ' \mvgee_{t-i}$ in \eqref{def-reducedrank_Z} to be approximated by Brownian motion in the limit as $d \to \infty$. The diminishing orders of $\mvV_i$'s suggest temporal independence of the VMA series in the limit $d\to\infty$, as are the increments of Brownian motion.

\section{VMA Representations}\label{s-VMA_rep}

We show here that both the series $\{\mvZ_t\}$ in \eqref{def-Z} and its reduced-dimension counterpart $\{\frac{1}{\sqrt{d}}\mvgL'\mvZ_t\}$ in \eqref{def-reducedrank_Z} have VMA representations. This is accomplished by appealing first to the Wold decomposition allowing one to write these series as VMAs of infinite order, with the underlying innovations having a special structure. We then argue that the matrix coefficients produced by the innovations algorithm converge to those in the Wold decomposition. Since the matrix coefficients of the innovations algorithm are exactly zero for covariances with finite support (as for $\{\mvZ_t\}$ from \eqref{ACVF_Z}), it will follow that the Wold decomposition needs to be a finite VMA. Details can be found in Section \ref{s-innovations}.

The considered innovations algorithm will thus provide a numerical means to compute the VMA representation in practice. Furthermore, the special cases of the innovations algorithm are exactly the algorithms considered in solving some non-linear matrix equations (NMEs), which are similar to \eqref{full dim Lag0}--\eqref{full dim Lag1-p}. We are unaware if this connection is known. It is explored in Section \ref{s-NME}.

\subsection{Innovations Algorithm and Existence of VMA Representations}\label{s-innovations}
Motivated by the discussion above, we first recall the innovations algorithm \cite{brockwell:2009}. The innovations algorithm is a recursive algorithm used to compute the best one-step-ahead linear predictor of a stationary time series given its ACVF. Suppose $\{\mvY_t\}$ is a $d$-dimensional zero mean stationary time series with the ACVF $\Gamma_Y(h) = \E \mvY_{t+h }\mvY_t'$ for $h \in \bZ$. Then, by the innovations algorithm, the one-step-ahead linear predictor $\wh\mvY_{n+1}$ and its prediction error covariance matrix $\mvgS^Y_n = \E (\mvY_{n+1} - \wh\mvY_{n+1})(\mvY_{n+1} - \wh\mvY_{n+1})'$ are given by 
\begin{equation}\label{def-onestepahead_pred}
\wh\mvY_{n+1} = \sum_{j=1}^n \mvgTh^Y_{n,j} (\mvY_{n+1-j} - \wh\mvY_{n+1-j}), \quad n \geq 1,
\end{equation}
with
\begin{align}
\mvgS_0^Y &= \Gamma_Y(0), \label{innov_1}\\
\mvgTh^Y_{n,n-k} &= \bigg ( \Gamma_Y(n-k) - \sum_{j=0}^{k-1} \mvgTh^Y_{n,n-j} \mvgS^Y_j (\mvgTh^{Y}_{k,k-j})' \bigg)(\mvgS^Y_k)^{-1}, \quad k=0,\ldots ,n-1, \label{innov_2}\\
\mvgS^Y_n &= \Gamma_Y(0) - \sum_{j=0}^{n-1} \mvgTh^Y_{n,n-j}\mvgS^Y_j (\mvgTh^{Y}_{n,n-j})'. \label{innov_3}
\end{align}
We will drop the superscript $Y$ when the dependence on the time series is clear from context. The terms $\mvY_{n+1-j} - \wh\mvY_{n+1-j}$ in \eqref{def-onestepahead_pred} are uncorrelated across $j$, and can therefore be viewed as innovations. Were $\{\mvY_n\}$ to have a VMA representation $\sum_{j=0}^{\infty} \mvgTh_j \mvgz_{n-j}$, one would expect (under the right conditions) for $\mvgTh_{n,j}$ to converge to $\mvgTh_j$. As noted above, this is the route we take in proving a VMA representation of $\{\mvZ_t\}$ in \eqref{def-Z}. An important observation in this regard is the following:
\begin{lem}\label{lem-innov_zeros}
$\Gamma_Y(h) = \mv0, \; |h| \geq p+1 \text{ implies } \mvgTh_{n,h} = \mv0, \; h \geq p+1$.
\end{lem}
This will be useful in deducing that the VMA representation of infinite order is in fact of finite order. The next theorem is the main result of this section. The proof following the above approach can be found in Appendix \ref{appendix-thm4.2}. The positive definite ordering on the set of real-valued $n \times n$ matrices is given by the relation $\mvM > \mvN$ if $ \mvM - \mvN$ is positive definite for $\mvM,\mvN \in \bR^{n\times n}$. A matrix $\mvM \in \bR^{n\times n}$ satisfying a condition $\cA$ is maximal with respect to positive definite ordering if for all $\mvN \in \bR^{n\times n}$ satisfying $\cA$, $\mvM > \mvN$.  

\begin{thm} \label{thm-VMArepZ}
Suppose $\{\mvX_t\}$ satisfies DFM \eqref{DFM}--\eqref{VARfactors} such that \eqref{error cov} and \eqref{lambda identity} hold. Then, the series $\{\mvZ_t\}$, as defined in \eqref{def-Z}, admits an invertible VMA$(p)$ representation 
\begin{equation}
\mvZ_t  = \mvgz_t + \sum_{i=1}^p \mvgTh_i \mvgz_{t-i},
\end{equation}
where $\{\mvgz_t\} \sim$ WN$(\mv0,\mvgS_{\gz})$ and $\mvgTh_i \in \bR^{d\times d}$ for $i = 1,\ldots,p$ satisfy the following properties with $\mvgS_{n},\mvgTh_{n,i}$ defined by the innovations algorithm \eqref{innov_1}--\eqref{innov_3} for $\{\Gamma_Z(h)\}_{h\in\bZ}$ : 
\begin{enumerate}
\item $\lim_{n\to \infty} \mvgS_n = \mvgS_{\gz} $,
\item $\lim_{n\to \infty} \mvgTh_{n,i} = \mvgTh_i \quad i= 1,\ldots, p$,
\item $\{\mvgS_{\gz},\mvgTh_1,\ldots,\mvgTh_p\}$ is the solution to \eqref{full dim Lag0}--\eqref{full dim Lag1-p} such that $\mvgS_{\gz}$ is maximal with respect to the positive definite ordering.
\end{enumerate}

Analogous results also apply to the reduced-dimension time series $\{\frac{1}{\sqrt{d}}\mvgL' \mvZ_t\}$ in \eqref{def-reducedrank_Z}.
\end{thm}

Theorem \ref{thm-VMArepZ} yields both the existence of a solution to \eqref{full dim Lag0}--\eqref{full dim Lag1-p} and a way to numerically compute a solution such that $\mvgS_{\gz}$ is maximal.

\subsection{Connections to Nonlinear Matrix Equations (NMEs)}\label{s-NME}
When $p=1$, the system of equations \eqref{full dim Lag0}--\eqref{full dim Lag1-p} becomes 
\begin{align}
\Gamma_Z(0) &= \mvgS_{\gz} + \mvgTh_1\mvgS_{\gz}\mvgTh_1', \label{Lag0EQ_ConnectionsNME} \\
\Gamma_Z(1) &= \mvgTh_1\mvgS_{\gz},\label{Lag1EQ_ConnectionsNME}
\end{align}
to be solved in terms of $\mvgS_{\gz},\mvgTh_1$. Solving \eqref{Lag1EQ_ConnectionsNME} for $\mvgTh_1 = \Gamma_Z(1)\mvgS_{\gz}^{-1}$ and substituting this into \eqref{Lag0EQ_ConnectionsNME} leads to a NME 
\begin{equation}\label{NME_p=1}
\Gamma_Z(0) = \mvgS_{\gz} + \Gamma_Z(1)\mvgS_{\gz}^{-1} \Gamma_Z(1)',
\end{equation}
as an equation in an unknown $\mvgS_{\gz}$ alone. The equation \eqref{NME_p=1} is of the NME form found in the matrix analysis literature \citep{anderson:1990, guo:1999, elsayed:2002} 
\begin{equation}\label{def-NME}
\mvX + \mvA' \mvX^{-1} \mvA = \mvQ,
\end{equation}
where $\mvX,\mvA,\mvQ \in \bR^{d \times d}$ and $\mvQ$ is positive definite. The matrices $\mvA,\mvQ$ are given and a solution $\mvX$ is sought. ($\mvX$ in \eqref{def-NME} and time series $\mvX_t$ in \eqref{DFM} should not be confused.) The NME \eqref{def-NME} arises in a diverse array of applications including control theory, fluid dynamics, stochastic filtering, dynamic programming and many more. The existence of solutions and numerical methods to compute such solutions have been studied extensively in matrix analysis literature (e.g.\ \citet{engwerda:1993}). 

It is interesting to look at the results concerning \eqref{def-NME} from a time series perspective. Consider, for example, the results of \citet{engwerda:1993} which classify all positive definite solutions to \eqref{def-NME}. Positive definite solutions of \eqref{def-NME} correspond to factorizations of the rational matrix-valued function 
\begin{equation}\label{def-psi}
\gPs(w) = \mvQ + w \mvA + w^{-1} \mvA', \quad w \in \bC,
\end{equation}
as the following result shows.
\begin{thm}{(\citet{engwerda:1993})}\label{thm-PDsolutions}
$\mvX$ is a positive definite solution to \eqref{def-NME} if and only if $\gPs$ is regular, i.e., there exists $w \in \bC$ such that $\det(\gPs(w)) \neq 0$, and for all $w$ on the complex unit circle, $\gPs(w)$ is semi-positive definite. 
If a positive definite solution exists, $\gPs$ factors as 
\begin{equation}\label{eq-factorpsi}
\gPs(w) = (\mvC_0' + w^{-1}\mvC_1')(\mvC_0 + w\mvC_1),
\end{equation}
where $\det(\mvC_0) \neq 0$ and $\mvX = \mvC_0'\mvC_0$ is a solution to \eqref{def-NME}. In fact, every positive definite solution is obtained this way. 
\end{thm}
In the case of NME \eqref{NME_p=1}, the function 
\begin{equation}
\gPs(w) = \gC_Z(0) + w\gC_Z(1) + w^{-1}\gC_Z(1)',
\end{equation}
with $w = e^{i\theta}$, is the spectral density of $\{\mvZ_t\}$ multiplied by $2\pi$. Theorem \ref{thm-PDsolutions} states that any solution $\mvgS_{\gz},\mvgTh_1$ must factorize the spectral density as 
\begin{align}\label{Z_spectraldensity}
\dfrac{\gPs(w)}{2\pi} &= \dfrac{1}{2\pi}( \mvgS_{\gz}^{1/2} + w^{-1}  \gC_Z(1)'\mvgS_{\gz}^{-1/2}) ( \mvgS_{\gz}^{1/2} + w \mvgS_{\gz}^{-1/2} \gC_Z(1)) \nonumber \\
&=  \dfrac{1}{2\pi}( \mvI_d + w^{-1}  \gC_Z(1)'\mvgS_{\gz}^{-1}) \mvgS_{\gz} ( \mvI_d + w \mvgS_{\gz}^{-1} \gC_Z(1)) \nonumber \\
&=  \dfrac{1}{2\pi}(\mvI_d + w^{-1} \mvgTh_1) \mvgS_{\gz} (\mvI_d + w \mvgTh_1').
\end{align}
Hence, through the lens of time series, Theorem \ref{thm-PDsolutions} states that the series $\{\mvZ_t\}$ admits a VMA representation if its spectral density factorizes as in \eqref{Z_spectraldensity}. Hence, Theorem \ref{thm-PDsolutions} is analogous to known results in time series research \cite{hannan:2012}. 

Furthermore, \citet{engwerda:1993} provide a recursive algorithm to numerically solve for a solution to a special case of \eqref{def-NME} when $\mvQ = \mvI_d$ and $\mvA$ is invertible. Note that solving the special case is in effect equivalent to solving the general case \cite{engwerda:1993}. The recursive algorithm proceeds as follows. Let $\mvS_0 = \mvI_d$. Then, update $\mvS_n$ by taking 
\begin{equation}\label{engwerda-num-sol}
\mvS_{n+1} = \mvI_d - \mvA'\mvS_n^{-1} \mvA,
\end{equation}
for $n \geq 1$. If \eqref{def-NME} has a solution, then \citet{engwerda:1993} show that $\mvS_n$ converges to $\mvS$, the maximal solution of \eqref{def-NME}. In other words, for any other solution $\wt\mvS$ to \eqref{def-NME}, $\mvS - \wt\mvS$ is positive definite. 

From a time series perspective, the recursive algorithm \eqref{engwerda-num-sol} is exactly the innovations algorithm applied to $\{\mvZ_t\}$ given $\gC_Z(0) = \mvI_d$. Indeed, when $p=1$, by Lemma \ref{lem-innov_zeros}, we know $\mvgTh_{n,h} = \mv0$ for all $h\geq 2$. So when $h = 1$, by \eqref{innov_2}, $\mvgTh_{n,1} =   \gC_Z(1)\mvgS_{n-1}^{-1}$. Thus, the innovations algorithm in the case of $p=1$ reduces to 
\begin{align}\label{def-S_n}
\mvgS_n &= \Gamma_Z(0) - \sum_{j=0}^{n-1} \mvgTh_{n,n-j}\mvgS_j \mvgTh'_{n,n-j} = \mvI_d - \gC_Z(1) \mvgS_{n-1}^{-1} \gC_Z(1)'.
\end{align}

Along with the recursive algorithm above, there exist many other algorithms to numerically solve for solutions of \eqref{def-NME}. For example, the following approach to numerically solve for a positive definite solution is presented in \citet{chiang:2016}. Let 
\begin{align}
\mvA^{(1)} & = \mvA \mvQ^{-1}\mvA, \nonumber \\
\mvB^{(1)} & = \mvA \mvQ^{-1} \mvA', \nonumber \\
\mvQ^{(1)} & = \mvQ - \mvA'\mvQ^{-1} \mvA, \nonumber \\
\mvQ^{(k)} & = \mvQ^{(1)} - (\mvA^{(1)})'(\mvQ^{(k-1)} - \mvB^{(1)})^{-1}\mvA^{(1)}, \;\; k \geq 2. \nonumber
\end{align} 
\begin{thm}{(\citet{chiang:2016})}\label{thm-numericalmaximalPDsolution}
Suppose there exists a positive definite matrix $\mvX_s$ such that
\begin{equation}\label{assump-numericalsolution}
\mvQ - \mvA'\mvX_s^{-1}\mvA - \mvX_s \geq 0 .
\end{equation}
Then, 
\begin{equation}\label{Q_infty}
\mvQ^{\infty} = \lim_{k \to \infty} \mvQ^{(k)}
\end{equation}
is the maximal positive definite solution of \eqref{def-NME}. 
\end{thm}
The condition \eqref{assump-numericalsolution} is shown in \citet{chiang:2016} to be equivalent to the required conditions of Theorem \ref{thm-PDsolutions}. The convergence \eqref{Q_infty} can be faster than that in the algorithm \eqref{def-S_n}, equivalent to the innovations algorithm.

However, when $p>1$, to the best of our knowledge, there are no results in the matrix analysis literature, theoretical or numerical, about the solutions of the system \eqref{full dim Lag0}--\eqref{full dim Lag1-p}. This may be due to the inability to reduce the case $p>1$ to a single matrix equation. For example, consider the case when $p=2$. Then, the system \eqref{full dim Lag0}--\eqref{full dim Lag1-p} becomes
\begin{align}
\gC_Z(0) &= \mvgS_{\gz} + \mvgTh_1 \mvgS_{\gz}\mvgTh_1' + \mvgTh_2 \mvgS_{\gz}\mvgTh_2',\label{p=2_acvf_1} \\ 
\gC_Z(1) &= \mvgTh_1 \mvgS_{\gz} \mvgTh_{0}' + \mvgTh_2 \mvgS_{\gz} \mvgTh_1', \label{p=2_acvf_2}\\
\gC_Z(2) &= \mvgTh_2\mvgS_{\gz}\mvgTh_0' \label{p=2_acvf_3}
\end{align}
where $\mvgTh_0 = \mvI_d$. Similar to the case $p=1$, we may solve for $\mvgTh_2 =  \gC_Z(2)\mvgS_{\gz}^{-1}$ and substituting $\mvgTh_2$ into $\gC_Z(0)$ and $\gC_Z(1)$ get 
\begin{align}
\gC_Z(0) &= \mvgS_{\gz} + \mvgTh_1 \mvgS_{\gz}\mvgTh_1' + \gC_Z(2) \mvgS_{\gz}^{-1} \gC_Z(2)',\label{p=2_reduced_gamma0} \\ 
\gC_Z(1) &= \mvgTh\mvgS_{\gz} + \gC_Z(2)\mvgTh_1'.\label{p=2_reduced_gamma1}
\end{align}
In the matrix analysis literature, \eqref{p=2_reduced_gamma0} is referred to as a Sylvester-Transpose equation \cite{hajarian:2013} when one knows $\mvgS_{\gz}$, $\gC_Z(1)$, and $\gC_Z(2)$ and aims for a solution for $\mvgTh_1$.  Although several results for the existence of a solution to a Sylvester-Transpose equation exist, we are not aware of any results in the matrix analysis literature concerning the systems of the type \eqref{p=2_reduced_gamma0}--\eqref{p=2_reduced_gamma1}. On the other hand, as shown above, we know that we can solve \eqref{p=2_acvf_1}--\eqref{p=2_acvf_3} or \eqref{p=2_reduced_gamma0}--\eqref{p=2_reduced_gamma1} through the innovations algorithm.

\section{VARMA Representations of DFMs}\label{s-main_res}

We gather here the various results obtained above in a single theorem concerning the DFM \eqref{DFM}--\eqref{VARfactors} (Section \ref{s-main_results}), and consider some of its implications (Sections \ref{s-forecasting} and \ref{s-reducedrank VAR}).
\subsection{Main Results}\label{s-main_results}

Collecting the results above, $\{\mvX_t\}$ following a DFM \eqref{DFM}--\eqref{VARfactors} can be rewritten as in \eqref{rewriteDFM} with the VAR$(p)$ component $\sum_{i=1}^p \wt\mvgF_i\mvX_{t-i}$. The remaining terms, by Theorem \ref{thm-VMArepZ}, have a VMA$(p)$ representation whose elements are defined by the maximal solution to the system \eqref{full dim Lag0}--\eqref{full dim Lag1-p}.  By the same theorem, so does $\{\frac{1}{\sqrt{d}}\mvgL'\mvZ_t\}$. Hence, the VMA$(p)$ representation of $\{\mvZ_t\}$ must admit a low-dimensional representation as given by Theorem \ref{thm-VMAdimred}. We summarize these results below.
\begin{thm}\label{DFMVARMA-thm}
	Suppose $\{\mvX_t\}$ satisfies DFM \eqref{DFM}--\eqref{VARfactors} such that \eqref{error cov} and \eqref{lambda identity} hold. Then:
	\begin{enumerate}
	\item $\{X_t\}$ admits a VARMA$(p,p)$ representation given by 
	\begin{equation}\label{VARMA-rep}
	\mvX_t = \sum_{i=1}^p \wt\mvgF_i \mvX_{t-i} + \mvgz_t + \sum_{i=1}^p \mvgTh_i \mvgz_{t-i},
	\end{equation}
with 
\begin{equation}\label{PhiTilde-mainresults}
\wt\mvgF_i = \frac{1}{d} \mvgL \mvgF_i\mvgL', \quad i=1,\ldots,p,
\end{equation}
$\{\mvgz_t\} \sim $ {\rm WN}$(\mv0,\mvgS_{\gz})$, $ \mvgS_{\gz} = \lim_{n\to \infty} \mvgS^Z_n $, $\mvgTh_i = \lim_{n\to\infty} \mvgTh^Z_{n,i}$, and $\mvgS^Z_n,\mvgTh^Z_{n,j}$ as in \eqref{innov_1}--\eqref{innov_3}. Also, $\{\mvgS_{\gz},\mvgTh_1,\ldots,\mvgTh_p\}$ is a maximal solution to \eqref{full dim Lag0}--\eqref{full dim Lag1-p} in the sense that $\mvgS_{\gz}$ is maximal with respect to the positive definite ordering. 
	\item Furthermore, there exist $\mvU,\mvV_i \in \bR^{r\times r}$, $i \in 1,\ldots,p$, with symmetric positive definite $\mvI_r+\mvU$ such that 
\begin{equation}\label{DFMVARMA-thm-lowdimrepres}
	\mvgS_{\gz} = \mvI_d + \dfrac{1}{d}\mvgL \mvU \mvgL' \text{,} \quad \mvgTh_i = \dfrac{1}{d}\mvgL \mvV_i \mvgL', \quad i= 1,\ldots ,p, 
\end{equation}
and the series $\{\frac{1}{\sqrt{d}}\mvgL'\mvZ_t\}$ admits a VMA$(p)$ representation with error covariance $\mvI_r + \mvU$ and VMA matrices $\mvV_1,\ldots,\mvV_p$. The matrices $\mvU,\mvV_i$ can be obtained by the innovations algorithm on $\{\frac{1}{\sqrt{d}} \mvgL'\mvZ_t\}$, i.e. \begin{equation}
\mvU =  \lim_{n\to \infty} \mvgS^{\frac{1}{\sqrt{d}}\mvgL'\mvZ_t}_n - \mvI_r, \quad \mvV_i = \lim_{n\to\infty} \mvgTh^{\frac{1}{\sqrt{d}}\mvgL'\mvZ_t}_{n,i}, \quad i =1,\ldots,p.
\end{equation}
\end{enumerate}
\end{thm}
\begin{rem} If $\mvgS_{\gee} \neq \mvI_d$, the VARMA matrices in \eqref{PhiTilde-mainresults} and \eqref{DFMVARMA-thm-lowdimrepres} are expressed using \eqref{PhiTilde_noniden_errorcov}. 
\end{rem}

\begin{rem} We note that the reduced rank structure $\mvgTh_i = \frac{1}{d}\mvgL \mvV_i \mvgL'$ for the VMA coefficients of $\mvZ_t$ and the fact that $\mvV_i$ are VMA coefficients of $\{\frac{1}{\sqrt{d}}\mvgL'\mvZ_t\}$ rely on the special form of $\mvZ_t$ in \eqref{def-Z}. In general, it is not true that for a VMA$(p)$ series $\mvZ_t = \mvgz_t + \sum_{i=1}^{p} \mvgTh_i \mvgz_{t-i}$, the VMA coefficients of the reduced-dimension series $\{\frac{1}{\sqrt{d}}\mvgL'\mvZ_t\}$ are necessarily of the form $d\mvgL'\mvgTh_i\mvgL$ as we are unable to write $\mvgL'\mvgTh_i\mvgz_{t-i} = \mvgL'\mvgTh_i\mvgL\mvgL'\mvgz_{t-i}$. 
\end{rem}

Note that $\mvgS_{\gz}$ is maximal with respect to the positive definite ordering which ensures the eigenvalues of $\mvU$ are as large as possible and based on numerical simulations it is often the case that $\mvU$ is positive definite. Furthermore, the reduced rank-structure of the DFM is a part of both the VAR and VMA components of \eqref{VARMA-rep} as stated in \eqref{PhiTilde-mainresults} and \eqref{DFMVARMA-thm-lowdimrepres}. The VAR matrices $\wt\mvgF_i$ and the VMA matrices $\mvgTh_i$ are of reduced-rank and $\mvgS_{\gz}$ is nearly of reduced-rank. This would lead one to believe that $\{\mvX_t\}$ is akin to a reduced-rank VAR model, but there exist important differences between the two models. A more thorough explanation is provided in Section \ref{s-reducedrank VAR}. As another implication of the approach leading to Theorem \ref{DFMVARMA-thm}, we can similarly deduce a low-dimensional structure in forecasting of the series $\{\mvX_t\}$.  

\subsection{Forecasting}\label{s-forecasting}
Theorem \ref{DFMVARMA-thm} allows one to use forecasting methods for VARMA models on the DFM $\{\mvX_t\}$ to compute the best linear predictors. Furthermore, as we show below, we may leverage the low-dimensional structure inherent in the VARMA representation of the DFM to calculate the predictors using significantly lower computational power. This can naturally be translated into an efficient likelihood calculation as well, though we do not pursue this line of investigation here. 

Consider the case when we wish to predict the one-step-ahead predictor of $\{\mvX_t\}$ given by 
\begin{equation}\label{def-onestepahead_pred_X}
\wh\mvX_{n+1} = \sum_{j=1}^n \mvgTh^X_{n,j} (\mvX_{n+1-j} - \wh\mvX_{n+1-j}), \quad n \geq 1,
\end{equation}
where $\mvgTh^X_{n,j}$ is given by \eqref{innov_1}--\eqref{innov_3}. We show that $\mvgTh^X_{n,j}$ has a similar representation as the VMA coefficients of $\mvZ_t$. This is due to the fact that the ACVF of $\mvX_t$ given by 
\begin{equation}\label{ACVF_X}
\Gamma_X(h) := \E \mvX_{t+h}\mvX_t' = 
    \begin{cases}
        \mvgL \E \mvf_{t}\mvf_t' \mvgL' + \mvI_d,  & \text{if } h = 0,\\
        \mvgL \E \mvf_{t+h}\mvf_t' \mvgL', & \text{if } h = 1,\ldots,p, \\
       \mv0, & \text{if } h \geq p+1
    \end{cases}
\end{equation}
has a similar structure to $\Gamma_Z(h)$ which was exploited in Theorem \ref{thm-VMAdimred}. More precisely, we have the following theorem whose proof is moved to Appendix \ref{appendix-thm5.4}. 
\begin{thm}\label{thm-innovs_reducedX}
Let $\mvW_t = \frac{1}{\sqrt{d}} \mvgL' \mvX_t$. Then,
\begin{align}
\mvgS^X_n &= \mvI_d + \frac{1}{d} \mvgL (\mvgS^W_n - \mvI_r) \mvgL', \label{innovs_reduced_sigma}\\
\mvgTh^X_{n,n-k} &= \frac{1}{d} \mvgL \mvgTh^W_{n,n-k} \mvgL', \quad k=0,\ldots ,n-1 , \label{innovs_reduced_theta}
\end{align}
for all $n \geq 1$.
\end{thm}

An immediate consequence of Theorem \ref{thm-innovs_reducedX} is the simplification of the one-step-ahead-predictor of $\mvX_t$. 
\begin{cor} We have
\begin{equation}
\wh\mvX_{n+1} = \sum_{j=1}^n \frac{1}{d} \mvgL \mvgTh^W_{n,n-k} \mvgL' (\mvX_{n+1-j} - \wh\mvX_{n+1-j}), \quad n \geq 1.
\end{equation}
\end{cor} 

\subsection{Connections to Reduced-Rank VAR Models}\label{s-reducedrank VAR}
Note again that the VAR transition matrices \eqref{PhiTilde-mainresults} in the VARMA representation \eqref{VARMA-rep} have a reduced rank. This may suggest that DFMs are akin to reduced-rank VAR models. The latter models and their applications have been studied by \citet{reinsel:1992} in the low-dimensional regime, and by \citet{basu:2019} and \citet{alquier:2020} in the high-dimensional regime, to name but a few. We shall argue here that there are in fact important differences between the two classes of models (DFM and reduced-rank VAR). 

To explain the differences, consider the following example. (The arguments apply more generally but we prefer to look at a special case for clarity.) Consider the DFM with one factor following an AR$(1)$ model as 
\begin{equation}\label{DFM-onefactor}
\begin{split}
\mvX_t &= \mv1 F_t + \mvgee_t, \\
F_t &= \phi F_{t-1} + \gh_t,
\end{split}
\end{equation}
where $\{\mvgee_t\}\sim$ WN$(\mv0,\mvI_d)$, $\{\gh_t\}\sim$ WN$(0,\gs^2_{\gh})$, and $\mv1$ is a $d\times 1$ vector of ones. As in Section \ref{s-rewriteDFM} and subsequent developments, we can write 
\begin{equation}\label{VARMA-onefactor}
\begin{split}
\mvX_t &= \phi \dfrac{\mv1\mv1'}{d}\mvX_{t-1} + \mv1 \gh_t + \mvgee_t - \phi\dfrac{\mv1\mv1'}{d}\mvgee_{t-1} \\
 &=: \phi \dfrac{\mv1\mv1'}{d}\mvX_{t-1} + \mve_t,
\end{split}
\end{equation}
where $\{\mve_t\}$ has a VMA$(1)$ structure. Now, consider also a reduced-rank VAR as a counterpart to \eqref{VARMA-onefactor} given by 
\begin{equation}\label{VARMA-onefactor-counterpart}
\mvY_t = \phi \dfrac{\mv1\mv1'}{d}\mvY_{t-1} + \mvga_t,
\end{equation}
where $\{\mvga_t\}\sim$ WN$(\mv0,\mvI_d)$. 

The models \eqref{VARMA-onefactor} and \eqref{VARMA-onefactor-counterpart} are different in the following ways. Note that \eqref{VARMA-onefactor-counterpart} can also be written in the form of \eqref{DFM-onefactor} as follows. Setting 
\begin{equation}\label{G_t}
G_t = \dfrac{\mv1'\mvY_{t-1}}{\sqrt{d}},
\end{equation}
we have 
\begin{equation}\label{DFM-Y_t}
\begin{split}
\mvY_t &= \phi \dfrac{\mv1}{\sqrt{d}} G_t + \mvga_t, \\
G_t &= \phi G_{t-1} + \gc_t,
\end{split}
\end{equation}
where $\{\gc_t = \mv1'\mvga_t / \sqrt{d}\}\sim$ WN$(0,1)$. The difference between \eqref{DFM-onefactor} and \eqref{DFM-Y_t} is in the loadings: $\mvgL = \mv1$ in \eqref{DFM-onefactor} and $\mvgL = \phi \mv1 / \sqrt{d}$ in \eqref{DFM-Y_t}. The former case is often referred to as that of strong factors and its theory is well-developed \citep{stock:2002, bai:2008, doz:2011}. The latter case is that of weak factors, with some theory available as well \citep{chamberlain:1982, uematsu:2019}; this case is arguably more difficult to deal with.

Another way to look at the difference between \eqref{VARMA-onefactor} and \eqref{VARMA-onefactor-counterpart} is that the error process $\{\mve_t\}$ and $\{\mvga_t\}$ have quite different properties: while $\{\mvga_t\}$ is a WN series, the series $\{\mve_t\}$ is constructed in a particular way. For example,
\begin{equation}
\dfrac{\mv1'\mve_t}{d} =  \gh_t + \dfrac{\mv1'\mvgee_t}{d} - \phi\dfrac{\mv1'\mvgee_{t-1}}{d} \to \gh_t  \quad \text{a.s.},
\end{equation}
whereas it is expected that 
\begin{equation}
\dfrac{\mv1'\mvga_t}{\sqrt{d}} \stackrel{d}{\to} N(0,1), \quad \dfrac{\mv1'\mvga_t}{d} \to 0 \quad \text{a.s.}
\end{equation}
This also results in different basic properties of the series $\{\mvX_t\}$ and $\{\mvY_t\}$ such as their correlation matrices. Indeed, straightforward calculations show that 
\begin{align}
\E \mvX_t \mvX_t'  &=\dfrac{\gs_{\gh}^2 }{1-\phi^2} \mv1\mv1' + \mvI_d, \\
\E \mvY_t\mvY_t' &= \dfrac{\phi^2}{1-\phi^2} \dfrac{\mv1\mv1'}{d} + \mvI_d,
\end{align}
and the respective correlation matrices are 
\begin{align}
\mvR_X &= \dfrac{\gs_{\gh}^2 }{\gs_{\gh}^2 + 1-\phi^2} \mv1\mv1' + \dfrac{1-\phi^2}{\gs_{\gh}^2 + 1-\phi^2} \mvI_d , \\
\mvR_Y &= \dfrac{\phi^2 }{\phi^2 + d(1-\phi^2)} \mv1\mv1' + \dfrac{d(1-\phi^2)}{\phi^2 + d(1-\phi^2)} \mvI_d .
\end{align}
Note that $\mvR_X$ has generally a more pronounced rank-1 component $\mv1\mv1'$ than $\mvR_Y$, except when $\phi$ is very close to 1. From a practical perspective, this also means that the sample correlation matrix $\wh\mvR_X$ will appear to have a rank-1 (or block) structure, whereas this structure will generally be ``hidden" in the case of $\mvR_Y$.  

\begin{rem}
The preceding discussion is also pertinent to DFMs and VAR models with network community structures as discussed in \citet{bhamidi:2022a}. 
\end{rem}

\section{Conclusions}\label{s-conclusion}

We have shown that a DFM \eqref{DFM} with factor series following a VAR$(p)$ model \eqref{VARfactors} can be represented as a VARMA$(p,p)$ model as given in \eqref{VARMA-rep}. The VAR$(p)$ component reveals itself by simply rewriting the DFM. After removing the VAR component from the DFM, we have shown that the left over component $\mvZ_t$ indeed admits a VMA$(p)$ representation. The existence of a VMA$(p)$ representation is shown by using the Wold decomposition (theorem) and the special structure of the DFM. In fact, by leveraging the low-dimensional structure of the DFM, we show that is enough to prove that the reduced-dimension counterpart $\{\frac{1}{\sqrt{d}} \mvgL' \mvZ_t\}$ admits a VMA representation. Hence, both the VAR and VMA components have a reduced-rank structure which we have explicitly given in \eqref{PhiTilde-mainresults}--\eqref{DFMVARMA-thm-lowdimrepres}. 

Furthermore, we have shown that the VMA components can be numerically calculated using the innovations algorithm, which solves a system of matrix equations given by the ACVFs. This system of non-linear matrix equations has been well studied in the matrix analysis literature for the case when $p=1$, and their results are interesting to look at from a time series perspective. In fact, one of the algorithms used to solve such a system in the matrix analysis literature is equivalent to the innovations algorithm when $p=1$. Lastly, we use the low-dimensional structure of the DFM to reduce computations of both the innovations algorithm and for forecasting. We have shown that for both $\{\mvZ_t\}$ and $\{\mvX_t\}$ it is enough to calculate the innovations algorithm for their reduced-dimension counterparts $\{\frac{1}{\sqrt{d}} \mvgL' \mvZ_t\}$ and $\{\frac{1}{\sqrt{d}} \mvgL' \mvX_t\}$, respectively. 

\appendix
\label{appendix}

\section{Proof of Theorem \ref{thm-VMAdimred}}\label{appendix-thm3.1}

In order to prove the statement, we show that the error covariance $\mvgS_{\gz}= \mvI_d + \frac{1}{d}\mvgL \mvU \mvgL'$ and  VMA matrices $\mvgTh_i = \frac{1}{d}\mvgL \mvV_i \mvgL'$ for $i = 1,\ldots, p$ satisfy the system \eqref{full dim Lag0}--\eqref{full dim Lag1-p} given that \eqref{lowdim-MVLag0}--\eqref{lowdim-MVLag1-p} hold. 

Rewriting the right-hand side of \eqref{full dim Lag0} and using \eqref{lambda identity}, we get 
\begin{align*}
\mvgS_{\gz} +  \sum_{i=1}^p\mvgTh_i \mvgS_{\gz} \mvgTh_i' &= \mvI_d + \frac{1}{d}\mvgL \mvU \mvgL' + \sum_{i=1}^p \frac{1}{d}\mvgL \mvV_i \mvgL'(\mvI_d + \frac{1}{d}\mvgL \mvU \mvgL')\frac{1}{d}\mvgL \mvV_i' \mvgL' \\ 
&= \mvI_d + \mvgL \bigg(\frac{1}{d}\mvU + \sum_{i=1}^p \frac{1}{d}\mvV_i (d\mvI_r + \frac{d^2}{d}\mvU )\frac{1}{d}\mvV_i' \bigg ) \mvgL' \\
& = \mvI_d + \mvgL \bigg( \mvgS_{\gh} + \frac{1}{d}\sum_{i=1}^p \mvgF_i\mvgF_i' \bigg ) \mvgL' \\ 
& = \mvgL\mvgS_{\gh}\mvgL' + \mvI_d +  \sum_{i=1}^p\wtmvgF_i \wtmvgF_i' ,
\end{align*}
where the second to last equality is obtained by \eqref{lowdim-MVLag0}. Hence, \eqref{full dim Lag0} holds. 

Similarly, rewriting the right-hand side of \eqref{full dim Lag1-p}, we obtain by \eqref{lowdim-MVLag1-p} that
\begin{align*}
\sum_{i=0}^{p-h} \mvgTh_i \mvgS_{\gz} \mvgTh_{i+h}'  & = \sum_{i=0}^{p-h} \frac{1}{d}\mvgL \mvV_i \mvgL' ( \mvI_d + \frac{1}{d}\mvgL \mvU \mvgL') \frac{1}{d}\mvgL \mvV_i' \mvgL' \\ 
& = \sum_{i=0}^{p-h} \frac{1}{d}\mvgL \mvV_i ( d\mvI_r +  \frac{d^2}{d}\mvU)\mvV_i' \mvgL'\frac{1}{d} \\ 
& = \sum_{i=0}^{p-h} \frac{1}{d} \mvgL \mvgF_i \mvgF_{i+h}\mvgL' = 
\sum_{i=0}^{p-h} \frac{1}{d^2} \mvgL \mvgF_i \mvgL' \mvgL \mvgF_{i+h}\mvgL' 
\\
&= \sum_{i=0}^{p-h} \wt\mvgF_i \wt\mvgF_{i+h},
\end{align*}
so \eqref{full dim Lag1-p} holds. 
$\blacksquare$

\section{Proofs of Lemma \ref{lem-innov_zeros} and Theorem \ref{thm-VMArepZ}}\label{appendix-thm4.2}

\textit{Proof of Lemma \ref{lem-innov_zeros}}: The innovations algorithm recursively obtains $\mvgTh_{n,j}$ in the following order: $\mvgS_0,\mvgTh_{1,1},\mvgS_1,\mvgTh_{2,2},\mvgTh_{2,1},\mvgS_2,\mvgTh_{3,3},\mvgTh_{3,2},\mvgTh_{3,1},$ etc. We proceed with the proof inductively on this ordering. If $h \geq p+1$, then $
\mvgTh_{h,h} = \Gamma_Y(h)\mvgS_0^{-1} = \mv0
$ by assumption. Let $n = h+1$. Suppose by induction that all $\mvgTh$'s computed before $\mvgTh_{n,h}$ in the innovations algorithm are zero. Then,
\begin{align*}
\mvgTh_{n,h} &= \bigg( \Gamma_Y(h) - \sum_{j=0}^{n-h-1} \mvgTh_{n,n-j} \mvgS_j \mvgTh'_{n-h,n-h-j} \bigg )\mvgS_{n-h}^{-1} \\
&= \bigg(-\sum_{j=0}^{n-h-1} \mvgTh_{n,n-j} \mvgS_j \mvgTh'_{n-h,n-h-j} \bigg )\mvgS_{n-h}^{-1} = \mv0,
\end{align*}
since $\mvgTh_{n,n-j} = \mv0$ for all $j=0,\ldots,n-h-1$. By induction, Lemma \ref{lem-innov_zeros} follows. $\blacksquare$

\bigskip

\textit{Proof of Theorem \ref{thm-VMArepZ}}:
We first show that $\{\mvZ_t\}$ has an infinite VMA representation. By Theorem 1.3.2 in \citet{hannan:2012}, it is enough to prove that the spectral density $f_Z$ of $\{\mvZ_t\}$ satisfies the condition 
\begin{equation}\label{lin-reg-req}
\int_{-\pi}^{\pi} \log(\det(f_Z(\theta))) d \theta > -\infty.
\end{equation}
By the definition of $\mvZ_t$ in \eqref{def-Z}, $f_Z$ can be written as the sum of spectral densities $f_{\gh}$ and $f_{\gee}$ of the two stationary series $\{\mvgL\mvgh_t\}$ and $\{\mvgee_t - \sum_{i=0}^p \wt\mvgF_i \mvgee_{t-i}\}$, respectively. Both $f_{\gh}$ and $f_{\gee}$ are positive semi-definite. Hence, by the super-additivity of determinants, 
\begin{align*}
\int_{-\pi}^{\pi}\log(\det(f_Z(\theta)))d \theta & \geq \int_{-\pi}^{\pi}\log(\det(f_{\gh}(\theta))+\det(f_{\gee}(\theta)))d \theta \\
&\geq \int_{-\pi}^{\pi}\log(\det(f_{\gee}(\theta)))d \theta > -\infty,
\end{align*}
where the last inequality follows from an application of Theorem 1.3.2 of \citet{hannan:2012} to the VMA series $\{\mvgee_t - \sum_{i=0}^p \wt\mvgF_i \mvgee_{t-i}\}$.  

Hence, by Theorem 1.3.2 of \citet{hannan:2012}, $\{\mvZ_t\}$ can be represented as an infinite VMA series. Note that there is no unique representation, but we may choose the representation using the linear innovations of $\mvZ_t$ as in the Wold Decomposition (Theorem 1.3.1 of \citet{hannan:2012}). We follow the notation and definitions of \citet{brockwell:2009}.  Define $\cM_{m}$ to be the Hilbert space in $\cL^2(\Omega)$ spanned by $\{\mvZ_t: t\leq m\}$ and $\cP_{\cM_m}$ be the projection onto $\cM_m$. Denote $\cP_{S}$ as the projection on to the space spanned by $\{\mvZ_t: t\in S\}$ for some set of integers $S$. Then, the VMA representation of $\{\mvZ_t\}$ in the Wold decomposition is given by 
\begin{equation}
\mvZ_t = \sum_{i = 0}^{\infty} \mvgTh_i \mvgz_{t-i},
\end{equation}
where $\mvgz_{t} = \mvZ_t - \cP_{\cM_{t-1}} \mvZ_{t}$, $\mvgTh_0 = \mvI_d$, and $\mvgTh_i = \E (\mvZ_t \mvgz_{t-i}') \mvgS_{\gz}^{-1}$, $i\geq 1$. The series $\{\mvgz_t\}$ is referred to as the linear innovations of $\{\mvZ_t\}$. 

In the innovations algorithm, the one-step-ahead predictor $\wh\mvZ_{n+1} = \cP_{1,\ldots,n}\mvZ_{n+1}$ and 
\begin{align*}
\mvV_n &:= \E ((\mvZ_{n+1} - \cP_{1,\ldots,n} \mvZ_{n+1})(\mvZ_{n+1} - \cP_{1,\ldots,n} \mvZ_{n+1})') \\ 
&= \E((\mvZ_{0} - \cP_{-n,\ldots,-1} \mvZ_0)(\mvZ_{0} - \cP_{-n,\ldots,-1} \mvZ_0)') \\
&\to \E((\mvZ_{0} - \cP_{\cM_{-1}} \mvZ_0)(\mvZ_{0} - \cP_{\cM_{-1}} \mvZ_0)') \\
&= \E \mvgz_0 \mvgz_0' = \mvgS_{\gz},
\end{align*}
where we use the fact that $\cP_{-n,\ldots,-1} \mvZ_0\to \cP_{\cM_{-1}}\mvZ_0$ in $\cL^2(\Omega)$ as given in problem 2.18 of \citet{brockwell:2009}. Similarly, 
\begin{align*}
\mvgTh_{n,i} &:= \E(\mvZ_{n+1}(\mvZ_{n+1-i} - \cP_{1,\ldots,n-i}\mvZ_{n+1-i})')\mvV_{n-i}' \\ 
&= \E(\mvZ_{i}(\mvZ_{0} - \cP_{-n-i,\ldots,-1}\mvZ_{0})')\mvV_{n-i}' \\
&\to \E(\mvZ_{i}(\mvZ_{0} - \cP_{\cM_{-1}}\mvZ_{0})')\mvgS_{\gz}' \\
&= \E(\mvZ_{i} \mvgz_0')\mvgS_{\gz}' = \mvgTh_i
\end{align*}
for $i \geq 1$. By \citet{rozanov:1967}, page 60, with this representation of $\mvZ_t$, $\mvgS_{\gz}$ must be maximal among all representations with respect to the positive definite ordering.

Since $\gC_Z(h) = \mv0, \; |h| \geq p+1 $, we have $\mvgTh_{n,h} = \mv0$ for all $h \geq p+1$ by Lemma \ref{lem-innov_zeros}. Then, $\{\mvZ_t\}$ has a VMA$(p)$ representation
\begin{equation}
\mvZ_t = \sum_{i = 0}^{p} \mvgTh_i \mvgz_{t-i}.
\end{equation}
$\blacksquare$

\section{Proof of Theorem \ref{thm-innovs_reducedX}}\label{appendix-thm5.4}
We first prove \eqref{innovs_reduced_theta}. The predictor is characterized by two properties. First, it is a linear function of the previous predictors and second, 
\begin{equation}\label{indep_requirement_proj}
\E (\mvY_{n+1} - \wh\mvY_{n+1})\mvY'_{n+1-i} = 1, \quad i = 1,\ldots,n.
\end{equation}
We will show that if $\mvgTh^W_{n,j}$ satisfies the conditions for $\wh\mvW_{n+1}$, as defined in \eqref{def-onestepahead_pred}, to be the best one-step-ahead linear predictor, then so does $\mvgTh^X_{n,j}$ as defined in \eqref{innovs_reduced_theta} for $\wh\mvX_{n+1}$. 

We do so inductively on $n$. When $n=1$, note that $\wh\mvW_2 = \mvgTh_{1,1}^W \mvW_1$. Then,
\begin{equation}\label{theta_base_case}
\begin{split}
\mv0 & = \E (\mvW_{2} - \wh\mvW_{2})\mvW'_{1} \\
&= \E (\mvW_{2} - \mvgTh_{1,1}^W \mvW_1)\mvW'_{1} \\
& = \E \bigg( \frac{\mvgL' \mvX_2}{\sqrt{d}} -  \frac{\mvgL'\mvgL}{d}\mvgTh_{1,1}^W\frac{\mvgL' \mvX_1}{\sqrt{d}} \bigg)  \frac{\mvX_1' \mvgL}{\sqrt{d}} \\
& = \frac{\mvgL'}{\sqrt{d}} \E \bigg[ \bigg (  \mvX_2 - \frac{\mvgL \mvgTh^W_{1,1} \mvgL'}{d} \mvX_1  \bigg )\mvX_1' \bigg ]  \frac{\mvgL }{\sqrt{d}}
\end{split}
\end{equation}
Note that $\E \mvX_2 \mvX_1' = \mvgL \E(\mvf_2\mvf_1') \mvgL' $ and $\E \mvX_1 \mvX_1' = \gC_X(0) =  \mvgL \E(\mvf_1\mvf_1') \mvgL' + \mvI_d$. So there exists $\mvA \in \bR^{r\times r}$ such that 
\begin{align*}
\E \bigg[ \bigg (  \mvX_2 - \frac{\mvgL \mvgTh^W_{1,1} \mvgL'}{d} \mvX_1  \bigg )\mvX_1' \bigg ] = \mvgL \mvA \mvgL'.
\end{align*}
By \eqref{theta_base_case}, $\mvA = \mv0$. Hence, $\mvgTh_{1,1}^X = \frac{1}{d}\mvgL \mvgTh^W_{1,1} \mvgL'$. 

Fix $n \geq 1 $ and suppose for $m \leq n$, $\mvgTh^X_{m,j} = \frac{1}{d}\mvgL \mvgTh^W_{m,j} \mvgL'$, $j = 1,\ldots,m$. Then, since $\wh\mvX_1 = \mv0$ and $\wh\mvW_1 = \mv0$,
\begin{equation}
\begin{split}
\wh\mvW_{2} &=  \mvgTh^W_{1,1} ( \mvW_{1} - \wh\mvW_{1})  \\ 
&  = \frac{\mvgL'}{\sqrt{d}} \bigg ( \frac{\mvgL \mvgTh^W_{1,1} \mvgL'}{d} (\mvX_1 - \wh\mvX_1) \bigg) \\
& = \frac{1}{\sqrt{d}} \mvgL'\wh\mvX_2.
\end{split}
\end{equation} 
If $\wh\mvW_{k} = \frac{1}{\sqrt{d}} \mvgL'\wh\mvX_k$ for $k \leq n$, then
\begin{equation}
\begin{split}
\wh\mvW_{k+1} &= \sum_{j=1}^k \mvgTh^W_{k,j} ( \mvW_{k+1-j} - \wh\mvW_{k-j}) \\
& = \dfrac{\mvgL'}{\sqrt{d}} \bigg ( \sum_{j=1}^k \frac{\mvgL \mvgTh^W_{k,j} \mvgL'}{d} (\mvX_{k+1-j} - \wh\mvX_{k+1-j}) \bigg ) \\
&= \dfrac{1}{\sqrt{d}} \mvgL'\wh\mvX_{k+1}.
\end{split}
\end{equation} 
So $\wh\mvW_{n+2-j}  = \dfrac{1}{\sqrt{d}} \mvgL'\wh\mvX_{n+2-j}$ for $j=1,\ldots,n+1$. Hence, for $i = 1,\ldots,n+1$,
\begin{equation}\label{theta_inductive_case}
\begin{split}
\mv0 &= \E (\mvW_{n+2} - \wh\mvW_{n+2})\mvW_i' \\
&= \E \bigg[ \bigg ( \mvW_{n+2} - \sum_{j=1}^{n+1} \mvgTh^W_{n+1,j} (\mvW_{n+2-j} - \wh\mvW_{n+2-j}) \bigg ) \mvW'_i \bigg ] \\
& = \dfrac{\mvgL'}{\sqrt{d}} \E \bigg [ \bigg ( \mvX_{n+2} - \sum_{j=1}^{n+1} \dfrac{\mvgL \mvgTh^W_{n+1,j}\mvgL'}{d} ( \mvX_{n+2-j} - \wh\mvX_{n+2-j} ) \bigg ) \mvX'_i \bigg ] \dfrac{\mvgL}{\sqrt{d}}.
\end{split}
\end{equation}
Note that $\E \mvX_{n+2} \mvX_i = \mvgL \E \mvf_{n+2}\mvf'_i \mvgL'$. If $i \neq n+2-j$, then $\E \mvX_{n+2-j} \mvX_i = \mvgL \E \mvf_{n+2-j}\mvf'_i \mvgL'$. If $i = n+2 - j$, then $\E \mvX_{n+2-j} \mvX_i =  \mvgL \E \mvf_{i}\mvf'_i \mvgL' + \mvI_d$. When $j = n+1$,  $\E \wh\mvX_{n+2-j}\mvX_i'  = \E \wh\mvX_{1}\mvX_i' = \mv0$. Suppose for $j = g + 1,\ldots,n+1$, there exists $\mvB_j \in \bR^{r \times r}$ such that $\E \wh\mvX_{n+2-j}\mvX_i' = \mvgL \mvB_j \mvgL'$. Then,
\begin{align*}
\E \wh\mvX_{n+2 - g}\mvX_i'  & = \E \sum_{ u= 1}^{n+1-g} \mvgTh^X_{n+1-g,u}( \mvX_{n+1-g-u} - \wh\mvX_{n+1-g-u})\mvX_i'  \\
& = \mvgL \mvB_g \mvgL
\end{align*}
for $\mvB_g \in \bR^{r \times r}$. So by induction, there exists $\mvB_j \in \bR^{r \times r}$, for $j = 1,\ldots,n+1$, such that $\E \wh\mvX_{n+2-j}\mvX_i' = \mvgL \mvB_j \mvgL'$. It follows that for some $\mvC \in \bR^{r \times r}$,
\begin{equation*}
\E \bigg [ \bigg ( \mvX_{n+2} - \sum_{j=1}^{n+1} \dfrac{\mvgL \mvgTh^W_{n+1,j}\mvgL'}{d} ( \mvX_{n+2-j} - \wh\mvX_{n+2-j} ) \bigg ) \mvX'_i \bigg ]  = \mvgL \mvC \mvgL'.
\end{equation*}
Hence, by \eqref{theta_inductive_case}, $\mvC = \mv0$ and 
\begin{equation}
\E \bigg [ \bigg ( \mvX_{n+2} - \sum_{j=1}^{n+1} \dfrac{\mvgL \mvgTh^W_{n+1,j}\mvgL'}{d} ( \mvX_{n+2-j} - \wh\mvX_{n+2-j} ) \bigg ) \mvX'_i \bigg ] = \mv0.
\end{equation}
Thus, by induction we have proven \eqref{innovs_reduced_theta}. 

To prove \eqref{innovs_reduced_sigma}, we use \eqref{innovs_reduced_theta} and \eqref{innov_1}--\eqref{innov_3}. For the case $n=0$,
\begin{align*}
\mvI_d + \dfrac{1}{d} \mvgL (\mvgS^W_0 - \mvI_r) \mvgL' &= \mvI_d + \dfrac{1}{d} \mvgL (\dfrac{1}{d} \mvgL' \mvgS_0^X \mvgL - \mvI_r) \mvgL' \\
&= \mvI_d + \dfrac{1}{d} \mvgL \bigg (\dfrac{1}{d} \mvgL' (\mvgL  (\E \mvf_0\mvf_0') \mvgL' + \mvI_d) \mvgL - \mvI_r \bigg ) \mvgL' \\
&= \mvI_d + \dfrac{1}{d} \mvgL(d (\E \mvf_0\mvf_0') + \mvI_r - \mvI_r) \mvgL'  \\
&= \mvI_d + \mvgL (\E \mvf_0\mvf_0') \mvgL'\\
& = \mvgS^X_0.
\end{align*}
For $n\geq 1$, by \eqref{innov_3},
\begin{align*}
\mvgS^X_n &= \gC_X(0) - \sum_{j=0}^{n-1} \mvgTh^X_{n,n-j} \mvgS^X_j (\mvgTh^X_{n,n-j})' \\
&= \mvI_d + \mvgL \mvI_r\mvgL' - \sum_{j=0}^{n-1} \dfrac{\mvgL \mvgTh^W_{n,n-j} \mvgL'}{d} \bigg (\mvI_d + \dfrac{1}{d}\mvgL(\mvgS^W_j - \mvI_r)\mvgL'\bigg) \dfrac{\mvgL (\mvgTh^W_{n,n-j})' \mvgL'}{d} \\ 
& = \mvI_d + \dfrac{\mvgL}{\sqrt{d}} \bigg [ d\mvI_r - \dfrac{1}{d}\sum_{j=0}^{n-1} \mvgTh^W_{n,n-j} \mvgL' \bigg (\mvI_d + \dfrac{1}{d}\mvgL(\mvgS^W_j - \mvI_r)\mvgL'\bigg)\mvgL (\mvgTh^W_{n,n-j})' \bigg ] \dfrac{\mvgL'}{\sqrt{d}} \\ 
&= \mvI_d + \dfrac{\mvgL}{\sqrt{d}} \bigg [ d\mvI_r - \sum_{j=0}^{n-1} \mvgTh^W_{n,n-j}\mvgS^W_j (\mvgTh^W_{n,n-j})' \bigg ] \dfrac{\mvgL'}{\sqrt{d}} \\
& = \mvI_d + \dfrac{1}{d} \mvgL (\mvgS^W_n - \mvI_r)\mvgL'.
\end{align*}
$\blacksquare$

\bibliographystyle{plain}
\bibliography{dfmvarma.bib}

\flushleft
\begin{tabular}{l}
Shankar Bhamidi, Dhruv Patel, Vladas Pipiras \\
Dept.\ of Statistics and Operations Research  \\
UNC at Chapel Hill \\
CB\#3260, Hanes Hall \\
Chapel Hill, NC 27599, USA \\
{\it bhamidi@email.unc.edu, dhruvpat@live.unc.edu, pipiras@email.unc.edu} \\
\end{tabular}

\end{document}